# Training for Open-Ended Drilling through a Virtual Reality Simulation

Hing Lie*
Wellesley College

Kachina Studer[†]
Mechanical Engineering, Massachusetts Institute of Technology

Zhen Zhao[‡]
Mechanical Engineering, Massachusetts Institute of Technology

Ben Thomson[§]
Mechanical Engineering, Massachusetts Institute of Technology

Dishita G Turakhia[¶]
CSAIL, Massachusetts Institute of Technology

John Liu[||]
Mechanical Engineering, Massachusetts Institute of Technology

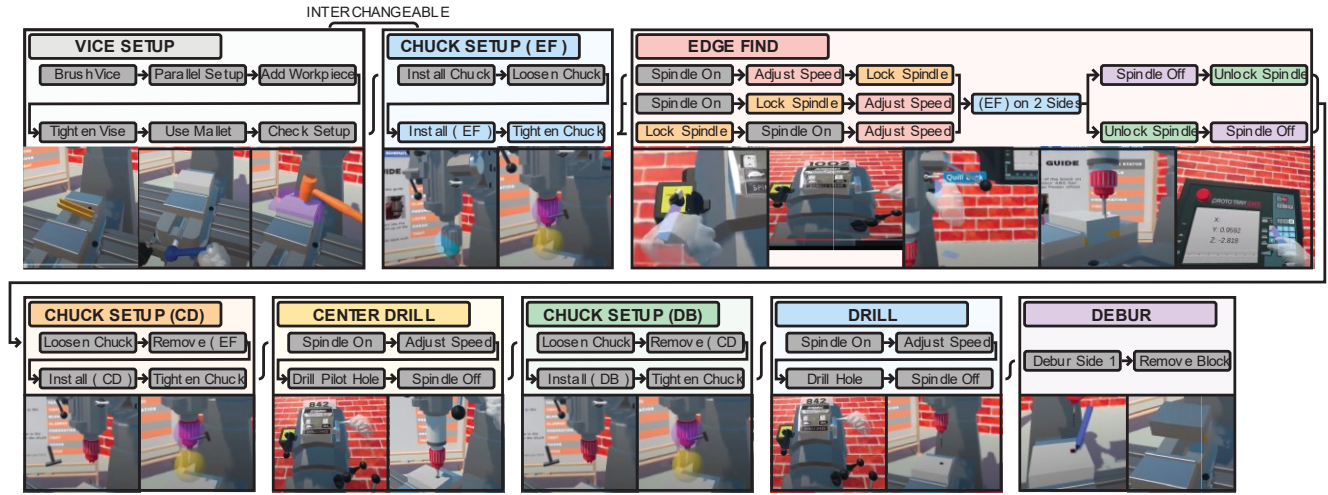

Figure 1: An open-ended VR system to train users to drill using a 3-axis milling machine. Our system allowed for multiple pathways to achieve the goal wherever possible and restricted the users to a single pathway wherever the tool required a strict protocol for operation. Arrows indicate pathways through tasks, a bracket indicates multiple pathways through tasks within one skill, with interchangeability between the tasks in vice set up and chuck set up.

## ABSTRACT

Virtual Reality (VR) can support effective and scalable training of psychomotor skills in manufacturing. However, many industry training modules offer experiences that are close-ended and do not allow for human error. We aim to address this gap in VR training tools for psychomotor skills training by exploring an open-ended approach to the system design. We designed a VR training simulation prototype to perform open-ended practice of drilling using a 3-axis milling machine. The simulation employs near "end-to-end" instruction through a safety module, a setup and drilling tutorial, open-ended practice complete with warnings of mistakes and failures, and a function to assess the geometries and locations of drilled holes against an engineering drawing. We developed and conducted a user study within an undergraduate-level introductory fabrication course to investigate the impact of open-ended VR practice on learning outcomes. Study results reveal positive trends, with the VR group successfully completing the machining task of drilling at a higher rate (75% vs 64%), with fewer mistakes (1.75 vs 2.14 score), and in less time (17.67 mins vs 21.57 mins) compared to the control group. We discuss our findings and limitations and implications for the design of open-ended VR training systems for learning psychomotor skills.

**Index Terms:** Human-centered computing—Interaction design—Interaction design process and methods—User centered design; Human-centered computing—Human computer interaction (HCI)—Interaction paradigms—Virtual reality

## 1 INTRODUCTION

Conventional training methods for psychomotor manufacturing skills, such as welding or machining, typically involve a combination of classroom instruction and psychomotor training [1], [2]. However, this approach is challenging to scale and distribute widely. Moreover, instructors spend a significant amount of time repeating instructions to new groups of trainees, leading to inefficiencies and high training costs [3]. The transfer of knowledge and skills within a region is also hindered by the speed at which people can learn from one another, particularly affecting rural areas' ability to upskill their workforce to adapt to advanced technology [4].

To address these issues, emerging technologies, particularly virtual reality (VR), offer potential solutions by providing scalable and personalized training for psychomotor skills in manufacturing [5]–[8]. Recent research has shown promising results in using VR for training machining tasks like welding, effectively preparing users for real work environments [9]. However, existing VR instructional systems for training often follow a step-by-step guided tutorial approach with limited room for open-endedness, user-engagement, or error management during training [10]. For instance, the VR training only offers the interactive of a button ("next") leaving no room for exploration [11]. In other instances where users are allowed relatively more freedom to interact with objects in VR, the system still restricts the goal completion to a single pathway [12]. This closed-loop approach limits the trainees' understanding of different tools and their overall training experience, as they miss out on learning multiple ways to achieve their goals.

*e-mail: hl3@wellesley.edu
[†]e-mail: kachinastuder@gmail.com
[‡]e-mail: zhenzhao@mit.edu
[§]e-mail: bt1@mit.edu
[¶]e-mail: dishita@mit.edu
[||]e-mail: johnhliu@mit.edu

This research aims to address this gap in the current work on VR training tools for psychomotor skill learning by exploring an open-ended approach to the system design. In particular, we designed an open-ended VR system to train users to drill using a 3-axis milling machine. Our system allowed for multiple pathways to achieve the goal wherever possible and restricted the users to a single pathway wherever the tool required a strict protocol for operation. For example, when setting up a milling machine for an edge finding operation, a user can lock the spindle, turn the spindle on, and adjust spindle speed, *in any order*, provided the spindle speed is adjusted *after* the spindle is on. Our open-ended system uses task-analysis and updates the learners' learning goals during training to allow them the freedom to explore multiple ways for task execution and goal adjustment.

To study the impact of our open-ended immersive VR training that adapts to human error during machining tasks on the learners' performance and training experience, we conducted a between-subjects user study with 26 university students enrolled in the introductory course Mechanical Engineering Tools at our institution. Study results reveal positive trends in favor of our open-ended VR system, with the VR training group successfully completing the machining task of drilling at a higher rate (75% vs 64%), with fewer mistakes (1.75 vs 2.14 score), and in lesser time (17.67 mins vs 21.57 mins) compared to the control group. Furthermore, the analysis of the participants feedback and our observation also provided insights into the psychomotor, cognitive, and affective aspects of skill-learning in VR.

This research paper makes the following contributions to the field of open-ended training systems in virtual reality (VR):

- A prototype of an open-ended VR training system for drilling using a 3-axis milling machine. The design of the prototype was based on the learning goals defined by instructors.
- A user-study to examine the impact of open-ended training on three key types of learning abilities: cognitive, affective, and psychomotor.
- A quantitative analysis of the study results and a qualitative analysis of the participants feedback that shed light on important implications for the design of open-ended VR training systems.

## 2 RELATED WORK

Several systems have been designed to allow exploration and open-endedness for learning both-knowledge based skills and motor skills.

### 2.1 Theories and Strategies for Open-ended Learning

In education, open-ended learning (OEL) and their facilitating environments (OELEs) promote the learner's determination of what is to be learned and how it is to be pursued [13]. In contrast with direct instruction (DI), which emphasizes the learner's ability to utilize prescribed learning strategies to achieve prescribed learning goals, OEL is heavily influenced by constructivism and emphasizes that understanding is mediated by the learner [14], [15]. OEL can enhance learning by providing flexibility [16], student-centered individualization [17], and self-directed learning [18].

In the field of motor skills, work has been carried out to explore learning strategies for the acquisition of skills. In particular, studies have examined the effect of practice in a fixed environment ("constant practice" - CP) versus practice in a changing environment ("variable practice" - VP) [19], [20]. In earlier studies, VP was found to be advantageous [21], but Ianovici recently demonstrated evidence that CP is more suited for novices, and VP more suited for experienced learners [22].

Researchers have explored the impact of instructional design based on constructivism on psychomotor acquisition. Cooper demonstrated that learners who have access to online instructional videos that can be paused and controlled by the learner can promote cognitive, affective and psychomotor rehabilitation skills [23]. Plummer demonstrated the perceived effectiveness of learning physical therapy in a virtual environment using a coaching model based on experiential learning theory and constructivism [24]. Of note, Melton found there was no difference in task success for learners who received instruction based on behaviorist versus constructivist theories, but the task at hand was linear with no alternate pathways (origami folding) [25].

Unlike the motor skill field which often studies one skill in isolation, manufacturing jobs require a wide-range of skills [26], a combination of fine and gross motor skills [5], and a discernment of the goal and situation to understand how to employ different skills to finish the job at hand. For example, a machinist must set up, operate, and disassemble machine tools. They must align, secure, and adjust cutting tools and workpieces. They will select different tools and machining specifications to turn, mill, drill, shape, and grind machine workpieces within given specifications [27]. A machinist may never fabricate the same part again. Yet little work has been carried out to incorporate theories and insights on open-ended learning from other fields to manufacturing skills training, such as learner set goals and strategies, and variable learning environments.

### 2.2 XR Systems for Learning Skills

XR systems have been successfully applied to the learning of skills within a variety of fields, including the medical, safety, engineering, and industrial fields [5]–[7]. In particular, VR appeals to psychomotor skills training because of the affordances of interactivity, embodiment, consistency, replicability, and safety [5]. XR has been found to assist the learning of gross and fine motor performance in assembly applications, surgical tasks, welding, and powered wheelchair maneuvering, and laboratory skills [5], [28].

In the training of motor skills, examples of open-ended training can be found in the development of intelligent tutoring systems (ITS). Here, learning goals are defined by the designer or coach (e.g. perform a squat, or shoot a ball), but there exist different ways to improve the user's performance such as multimodal feedback, actuation for motor task correction, and actuation for varying motor task difficulty [29]. Developments have also explored how systems should instruct or recommend real-time adjustments to users [30].

Within manufacturing skills training, much of the research in extended reality (XR) training has focused on assembly and welding skills [5], [8]. Chan categorized XR affordances that promote psychomotor skills in welding as training benefits (gestures, multimodal and sensory interactions, etc.), post-training assessment (performance review, playback, etc.), and instructor assistance (analytics for instructors, real-time guidance) [8]. Other reviews have classified XR training by industrial task type [31], target population [32], device type [31], development platforms [33], and challenges and learning limitations [32].

While there is work grounded in constructivism to develop VR learning environments, prior work has focused on cognitive skills [34]–[36]. To our knowledge there is limited work classifying virtual environments by their open-endedness or exploring their impact on psychomotor skills.

## 3 SYSTEMS DESIGN AND DEVELOPMENT

We applied a backward design approach from the learning sciences [37] to identify learning outcomes for a drilling operation. These learning outcomes were broken down into the necessary tasks students need to be able to accomplish. We then analyzed these tasks to identify challenging concepts and the most common errors for novice learners. These inputs informed the design and development of a virtual reality training simulation designed to provide students the opportunity to train and practice in an open-ended environment.

### 3.1 Learning Outcomes and Task Analysis

We collaborated with five machining instructors to define the learning outcomes of the training activity and carry out task analysis. Three instructors who have taught machining in a university-setting for at least five years. Two instructors are paper authors who have spent hundreds of hours of training students on machining.

From our discussion with machining instructors, we identified the learning outcomes for using a manual mill to carry out the basic operation of drilling (Table 1). More specifically, the goal was for students to properly set up a mill to drill a hole in a block of material within a given geometric tolerance. We chose to focus on drilling because it was simple compared to other machining operations such as milling or turning, but still had many pathways to successfully complete the goal. We broke these learning outcomes into individual tasks. For example, to properly set up the workpiece using the vise and parallels (Learning Outcome #2), a user should be able to use a brush to clear the vise from any remaining metal chips, add precision machined parallels to ensure the surface of the workpiece is flat with respect to the horizontal plane of the machine, loosen and tighten the vise, hit the workpiece with a mallet to ensure the workpiece is snug against the parallels, and to check whether the parallels can slide in and out or are snug.

Table 1: Learning outcomes and summary of associated tasks.

| Learning Outcome | Summary of Tasks |
| --- | --- |
| Choose proper clothing and attire to operate safely in the machine shop. | • Put on safety goggles<br>• Fix attire or physical attributes that present safety hazards (tie hair up, take rings off, etc.)<br>• Avoid touching active tooling |
| Set up the workpiece using the vise and parallels. | • Clean vise and secure part<br>• Add parallels to properly support the material<br>• Tighten vise and hit part with a mallet to ensure flatness |
| Set up the mill's coordinate system by operating the table handles, edge finder, and DRO. | • Use an edge-finding tool to reference the milling machine's coordinate system with the part's dimensions<br>• Gently guide to the edge of the part and notice exactly when it has made contact |
| Use the mill and tools to center drill, drill, and deburr, including pecking and setting the speed of the spindle. | • Remove the previous tool from the machine's chuck and install the appropriate tool<br>• Use the machine's handwheels to position the center of the tool at a specified location<br>• Operate drilling within the correct ranges of spindle speed<br>• Operate pecking during drilling operations when necessary |

For each task, instructors identified the errors novice learners most commonly make (Appendix). These could range from i) actions not aligned with best practice (e.g., used mallet lightly), ii) actions that risk the quality of the part or health of the tool or machine (e.g., did not debur or peck), or iii) actions that presented risk to the operator (e.g., brushed down vise with hand, did not tie hair.) We also assessed that some of these errors commonly occur because those tasks heavily depend on the senses for feedback.

For example:

- ***Pecking*** is the action of driving the rotating drill bit into the material a small distance before retracting it, then repeating with successively larger depths. If the user simply plunges the drill bit directly into the work, the metal chips cannot vacate. This can lead to a stuck or broken tool, excessive heat generation, a misshapen hole, and poor surface finish. Experts may augment their sense of the distance traveled with the physical sense of resistance, while novices can struggle to tell if the drill bit is going too far or not far enough.
- ***Machining parameters****:* Feed defines the rate at which the tool is moved into the workpiece. Speed is the rotational speed of that tool. Proper setting of these parameters heavily relies on the tool diameter. Together, these settings will also affect part quality and tool life. An experienced machinist relies on visual cues (noticing differences in chip formation) and audio cues (the sound of rotating spindles or cutting action) to double-check whether the feeds and speeds are correct.

### 3.2 VR Design Principles

Borrowing from [9], application features for learning guides, machine environment, and tools were carefully considered to support perceptual, cognitive, motor development for participants (Table 2). Three spatial elements were devised to provide clear structure with the intent to support task awareness, skill development, and workflow.

### 3.3 VR Simulation

We developed the system for the Meta Quest 2 VR stand alone head mounted display (HMD) with two Quest 2 controllers, using the Unity3D game development environment (2020.3.13f). The Oculus XR Plugin was used for setting up deployment of the APK from Unity for the Quest headset. The XR Interaction Toolkit 2.0.3 was used for integrating the inputs and object interactions. Our code is available open-source [38].

The system allows users to interact with any tool, workpiece, machine component, button, or switch at any time. The system will warn the user of performance errors they make (Appendix), such as using the center drill to drill too deep or using an incorrect spindle speed. During open-ended practice, the system allows users to perform any action within the multiple allowable pathways set by task mapping (Figure 1). For each attempted action, the system checks the current state of the machine set up and the prerequisite tasks that need to be completed before this action is allowable. If a user attempts a task that deviates from allowable pathways, the machine will display an error message on the whiteboard. If the attempted task would cause harm to the person or tool in the physical context, the machine will stop, and give users the opportunity to accept the error message to continue. During the drilling module with embedded instruction, the system allows open interactions in the same way as open-ended practice. However, the system gates each instructional step to support users completing drilling using one correct task order.

The system facilitates interactions using two controller buttons: grip and trigger. Users use the grip button to grab and hold tools, workpieces, machine components (levers or table wheels), and area teleportation. Users use the trigger button to press buttons or switches on the machine or whiteboard. These controller interactions also support extended reach; for example, if a machine button is too far away for reach, the system provides a line pointer to press the button.

**Milling Machine** (Figure 2a): The milling machine model was based off of a Bridgeport EZ-TRAK DX 3-axis manual mill [39]. The machine components and interactions were developed for authenticity and layered with visual cues to provide additional guidance. The active components of the modeled machine include: knee, column, 2-axis table, quill, spindle, vise, and digital readout.

Table 2: VR design principles.

| MICROSKILLS | LEARNING ELEMENTS | VR DESIGN | VR EXAMPLES |
|---|---|---|---|
| Perceptual | • Establish workflow of physical machine environment<br>• Abstracted visual indicators for interaction points<br>• Audio feedback from machining interactions and operation | • Visually distinct areas of interactions [A]<br>• Machining snap zones [B]<br>• Interactions with audio feedback [C] | [A] [B] [C] |
| Cognitive | • Explain assignment through learning category / tasks<br>• Teach Tool & Machine Mechanics<br>• Teach machining process<br>• Explain abstract concepts | • Adaptable Step by step written guide and diagrams for machining tasks [D]<br>• Video tutorials page with section guide ( play / pause / advance ) [E]<br>• Instructional Review PDF [F]<br>• Warning and failure states [G]<br>• Drilling Schematic & Data [H] | [D] [E] [F] [G] [H] |
| Motor | • Interactive guide for user interaction types<br>• Provide precise feedback for fine motor interactions<br>• Provide accuracy for fine motor interactions | • Clear instruction for controller interactions [I]<br>• Visual feedback for precise machine interaction [J]<br>• Accuracy of movement for machine precise interactAions [K] | [I] [J] [K] |

**Whiteboard** (Figure 2b): Instructional guide tabs were provided as step-by-step tasks, video modules, and PDF review guide. A workpiece schematic including locations of any drilled holes was also provided for self-assessment. Warning and errors were displayed on the whiteboard when students incorrectly performed tasks. These were dismissible, allowing students to continue to practice.

**Tool Board** (Figure 2c): The board of tools was placed in a familiar location and with tool labels to reduce users' cognitive load. Tools were snappable and removable from the board allowing participants to complete tasks and for tools to return to an exact location. These tools included the chuck, chuck key, mallet, edge finder, center drill, parallels, debur tool, and a range of drills of varying diameters.

The simulation was designed to replicate the physical environment of a student machine shop: a 3-axis milling machine, tools, and written and video instructions (Figure 2).

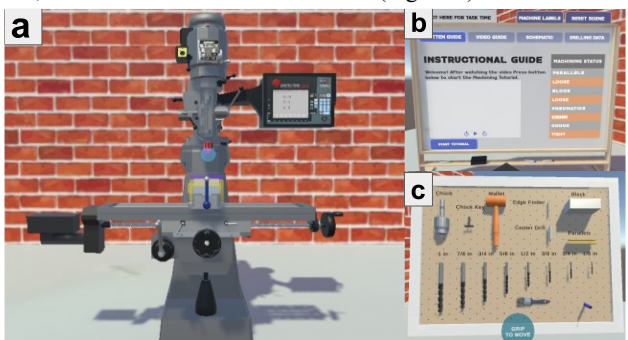

Figure 2: The VR simulation is composed of a a) three-axis manual mill, b) a white board of instructions and guidance, and c) a board full of tools.

*Instruction:* Users receive instruction through three formats:
1. A brief video introduction was provided to familiarize participants with the instructor, followed by step-by-step instructions and diagrams which advanced automatically upon completion.
2. A PDF guide that was presented to control group participants was also provided to students for review at the end of the VR training simulation.
3. Supplementary instructional elements were also integrated as part of the user interface. Labels for machine parts and tools were added to enhance familiarity within the machine and tool set for the assessment.

*Feedback*: most of the feedback during the activity was given to students through warning and error messages on the whiteboard. We also made unique developments to help students learn concepts for which experts rely heavily on the senses for feedback.

To help students learn proper pecking during drilling, we developed a "heat generation" simulator. When the tool is engaged in the workpiece, it accumulates heat as a function of its speed, feed rate, and diameter. When the drill is moved up to clear the work (pecking), the system reduces the tool's stored "heat" value. By properly pecking the tool, users can ensure that the heat value never exceeds a predetermined threshold. If the heat value does exceed that threshold, an error sign appears to alert students they have not properly employed pecking when drilling.

When actively cutting, the simulation also varies the sound of the cutting based on the feed and speed of the tool. We produced these sounds by recording the actual sounds of drilling using the manual mill in our machine shop. When the user is operating the mill within its proper ranges, a quieter recording is played. If the tool is moved too aggressively into the work, a grinding sound indicates that something is not right. If the spindle speed increases, the pitch of the sound of spinning increases as well.

### 3.4 Learning Modules

Our VR training simulation was divided into four modules. Users become familiar with VR interactions, learn about safety, learn about drilling through an interactive tutorial, and then practice in an open-ended environment (Figure 3).

**VR Tutorial:** The VR tutorial teaches the user how to use a VR controller to perform basic operations in the simulation. It shows the user how to grab or throw objects, and push buttons. The tutorial takes a few minutes to complete, and provides users with the VR knowledge they need to train in subsequent modules.

**Lab Safety Module:** The Safety Module takes the user into a space with a whiteboard. Users operate on the two-dimensional surface of the whiteboard, which helps further integrate them into the virtual environment. This module takes a few minutes to complete, and addresses the safety aspect of machining, which is vital but often overlooked by VR machining training simulations. During this module, users are tasked with preparing their avatar, displayed on the whiteboard, for the laboratory environment. This involves identifying and fixing any potential safety hazards, such as removing jewelry and putting on goggles. Once the avatar is prepared for the lab, the user can proceed to the next module, as if they are now safely entering into the machine shop.

**Drilling module with embedded instructions:** Within the drilling module, users enter into a space with a whiteboard, a milling machine, and a tool board. They learn how to carry out operations to drill a hole into a 4"x6" metal block. Students watch short video tutorials recorded by a machining instructor. Students

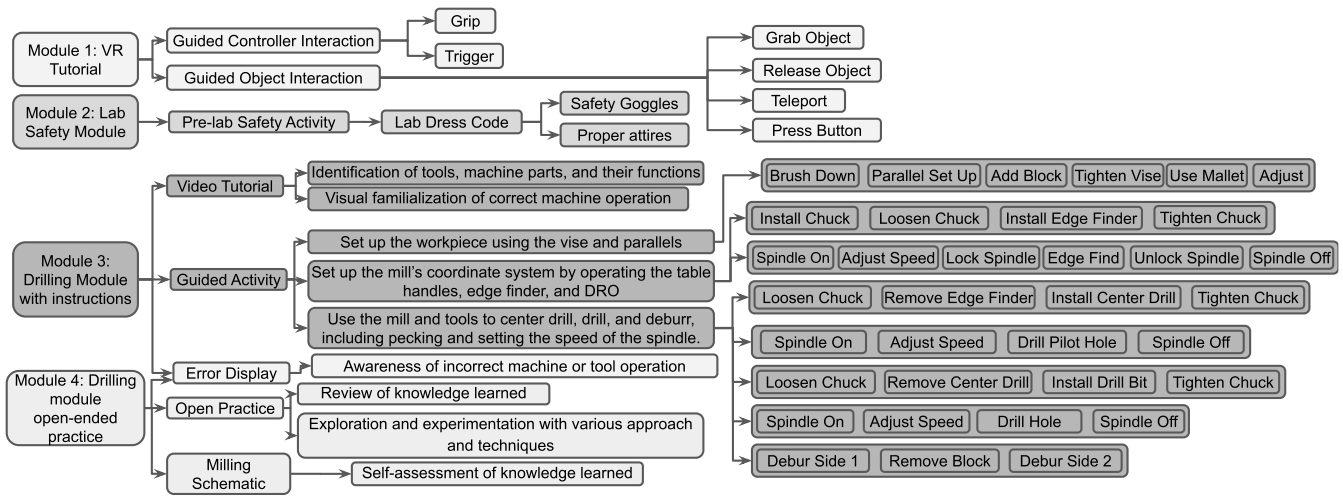
Figure 3: Flow of learning modules: VR tutorial, lab safety, drilling with instructions, and open-ended drilling practice.

are guided by step-by-step instructions that automatically update according to the user's progress. The guide does not gate a user from what they can and cannot do, but rather guides them on one of the correct paths to successfully completing the drilling task.

**Drilling module open-ended practice:** This module allows users to practice what they have learned in an open-ended setting. Users can self-assess their skills using tools including a schematic that also displays the position of each hole they drill.

### 3.5 Simulation Open-endedness

Training systems can be closed-ended when they i) prevent a user from interacting with objects not directly associated with a prescribed task, or ii) do not advance instruction until a prescribed task is completed – even when in reality multiple task orders can complete the goal equally well.

Our system allowed for multiple pathways to achieve the goal of drilling wherever possible (Figure 1). For example, when setting up a milling machine for an edge finding operation, a user can lock the spindle, turn the spindle on, and adjust spindle speed, *in any order*, provided the spindle speed is adjusted *after* the spindle is on.

However, when there is only one order to operate a set of tasks correctly, the simulation requires this one order for users to succeed. For instance, when the drill bit has been installed into the chuck, the succeeding action of drilling a hole has only one correct task order. The user must turn on the spindle, adjust the spindle speed, operate the quill handle to drill the hole, and then turn off the spindle.

Finally, the simulation does not gate tasks within one skill from another when the real-life task would not require it. For example, a user in the simulation can also succeed if they set up the chuck and vise simultaneously instead of only one after another.

## 4 USER TESTING AND EVALUATION

This section describes the study set up, participants, study procedure, and results and analysis.

### 4.1 Study Set Up

The goal of this study was to evaluate the VR-based machining tutorial system within cognitive, affective, and psychomotor domains. The training performance was evaluated using five assessments:
- Assessment 1: a survey to measure attitudes
- Assessment 2: a quiz to measure cognitive understanding
- Assessment 3: a drilling task evaluation
- Assessment 4: the NASA task load index (NASA-TLX) to measure workload
- Assessment 5: a subjective questionnaire to obtain general feedback on students' learning experience

### 4.2 Participants

The study was conducted with 26 students enrolled in the introductory course Mechanical Engineering Tools and were primarily students intending to pursue a major in mechanical engineering. We illustrate their class standing and prior machining experience in Table 3. Students were assigned into control and VR groups to match their schedule preferences. Instructors also tried to maintain a similar level of overall machining experience between the two groups.

Table 3: Learner class standing and prior machining experience.

|  | Freshman | 7 |
|---|---|---|
| Class Standing | Sophomore | 17 |
|  | Junior | 2 |
|  | Senior | 0 |
|  | 0 | 3 |
| Prior machining experience (hours) | 1-10 | 15 |
|  | 11-50 | 6 |
|  | 51-200 | 1 |
|  | 200+ | 1 |

### 4.3 Study Procedure

Figure 4 shows the flow chart of the user test. Students first carried out a pre-assessment activity composed of affective and cognitive assessments (Assessments 1 and 2). The affective assessment consisted of 21 questions on a 7-point Likert scale (1- "not at all true of me" to 7 – "very true of me") adapted from the Motivated Strategies for Learning Questionnaire (MSLQ) [40]. The affective assessment measured three constructs of self-efficacy, namely the student's belief in one's capacity to: understand the content, use the equipment, and perform in the class. The assessment also measured students' motivation to re-engage with the content and their fear of making mistakes. The cognitive assessments consisted of 21 multiple-choice questions on technical aspects of drilling. Some questions were adapted from the National Institute for Metalworking Skills (NIMS) Machining Level 1 Preparation Guides for Milling [41]. These technical questions covered component identification and purpose, safety, and set up and operations.

The drilling tutorial was delivered to the VR group in the VR-based environment, and to the control group in a traditional classroom. Instructors and researchers were present to help them situate themselves with the VR module, but did not answer questions on the training material. Students in both groups were given 50 minutes to finish the training segment.

After the training segment, both groups were given 10 minutes to carry out a post-assessment activity composed of the same questions from the pre-assessment (Assessment 1 and 2).

Students were then brought into the machine shop and given 30 minutes to use a manual mill to carry out a drilling operation (Assessment 3). They were given a set of manual tools and a blueprint with the location and dimensions of the intended hole. Instructors were present and could answer questions during the drilling task. Researchers were also present to record time on task, errors in real-time, and any questions that were asked during this session. Errors in real-time were then converted into a weighted score. Errors' weight (0 to 4) reflects the potential risk incurred by the errors as assessed by instructors. Each participant received a total error score by summing the weighted score of all the errors that person committed. A higher score indicates more severe errors committed in total.

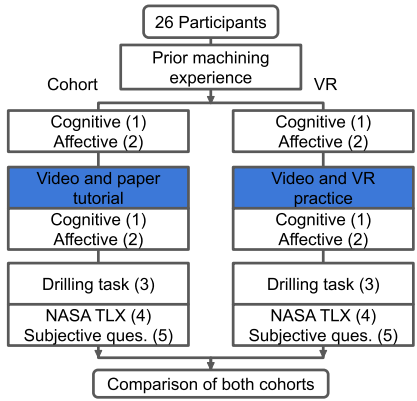

Figure 4: Flowchart of the user test. The goal was to compare learning gains between control and VR-based machining training within cognitive, affective, and psychomotor domains.

Students were then given the *NASA Task Load Index* questionnaire (NASA-TLX) [42] to measure workload assessment (Assessment 4) and a subjective questionnaire for students to reflect on their experience and give feedback (Assessment 5).

### 4.4 Results and Analysis

We examined affective, cognitive, and psychomotor outcomes, in addition to perceived workload and learning experience. Each analysis centered on the mean comparisons between the VR and control groups. All analyses were conducted using IBM SPSS (28.0.1.1).

Little's MCAR test [43] indicated the data is missing completely at random (p=.88). While the missing rate is 1% < 5%, the missing data were omitted pairwise during the analyses. The sample sizes for the VR group and the control group are 12 and 14, respectively.

To counter presurvey score bias, we adopted individual normalized change [44] to compare the mean score change of both affective and cognitive constructs between the VR and control groups. Participants' individual normalized change (c) is calculated as:

$$c = \begin{cases} \dfrac{postscore\% - prescore\%}{100\% - prescore\%}, & if\ postscore > prescore \\ \dfrac{postscore\% - prescore\%}{prescore\%}, & if\ postscore < prescore \end{cases}$$

This metric measures the pre/post-changes of each individual participant. If a participant is scored at 100% for both pre- and post-assessment, it is not possible to measure a change regardless of the intervention type. Individual normalized change takes this into account by excluding such pair-wise data.

We constructed Shapiro-Wilk tests and Q-Q plots to examine the normality of the data (p<.05). The results indicated that some measures were normally distributed for both VR and control groups and some were not (Table 4).

Table 4: Breakdown of normally and not normally distributed measures.

|  | Measures for which normality assumption held for both groups | Measures for which the normality assumption did not hold for both groups |
|---|---|---|
| Affective | Self-efficacy (use), Self-efficacy (performance), Fear of mistakes | Self-efficacy (understand), Motivation to re-engage |
| Cognitive | Component identification and purpose, Set up and operations, Total cognitive score | Safety |
| Psychomotor | Time on Task | Total error score, No. of questions |
| NASA TLX | Mental Demand, Physical Demand, Effort, Frustration | Temporal Demand, Performance |
| Learning experience | N/A | Reflection 1, Reflection 2 |

We used both parametric and nonparametric tests to analyze the data due to the coexistence of normality and non-normality [45], [46]. We used T-tests to compare the group means over the measures for which data in both groups are normally distributed, and Mann-Whitney-U tests to compare group distributions over measures for which data in the two groups are not both normally distributed. For all comparisons, a positive value indicates that the mean value of the VR group was higher than that of the control group. A negative value means the mean value of the VR group was lower than that of the control group.

Table 5 compares the average cognitive gains between the groups. While these findings are not statistically significant, the VR group scored larger gains both overall and in each category (component identification, safety, set up and operations), which indicates a positive trend.

Table 5: Cognitive assessments: VR vs control groups.

|  | Total cognitive scores | Component identification and purpose | Setup and operations |  | Safety |
|---|---|---|---|---|---|
| T stat | .84 | .72 | 1.00 | U stat | 11.5 |
| p-value | .41 | .48 | .33 | p-value | .65 |

Table 6 compares the average affective gains between the VR and control groups. With the exception of the self-efficacy to use equipment where the VR group scored a larger gain, the VR group scored smaller gains than the control group on all other affective measures. In particular, the smaller gain for self-efficacy on understanding course material and content was statistically significant (p<0.01).

Table 6: Affective assessments: VR vs control groups.

|  | Self-efficacy | | Fear of mistakes |  | Self-efficacy (understand) | Motivation to re-engage |
|---|---|---|---|---|---|---|
|  | Use | Perform |  |  |  |  |
| T stat | .76 | -1.15 | -1.24 | U stat | 27 | 39.5 |
| p-value | .45 | .89 | .23 | p-value | .02* | .07 |

* p-value <0.05

Table 7 compares the mean values of psychomotor assessments between the VR and control groups. While these findings are not statistically significant, the VR group scored lower (better) in all constructs: participants' weighted total error score, time spent on drilling tasks, and number of questions asked, indicating a positive trend. Further, 75% of the VR group successfully finished drilling a hole within the specified tolerance, as opposed to 64% of the control group.

Table 7: Psychomotor assessments: VR vs control groups.

|  | Time on Task | | Total error score | No. of questions | | Drilled Block Measurement |
|---|---|---|---|---|---|---|
| T stat | -1.54 | U-stat | 73.5 | 72 | Control group | 64% |
| p-value | .14 | p-value | .60 | .56 | VR group | 75% |

Table 8 compares the mean of the NASA TLX measures between VR and control groups. While not statistically significant, the VR group reported less time pressure for the drilling task. However, they reported exerting more mental acuity, physical activity, and effort.

Table 8: NASA TLX: VR vs control groups.

|  | Mental Demand | Physical Demand | Effort | Frustration |  | Temporal Demand | Performance |
|---|---|---|---|---|---|---|---|
| T stat | .84 | .31 | .26 | -.01 | U stat | 67.5 | 81 |
| p-value | .41 | .76 | .80 | .99 | p-value | .40 | .85 |

Table 9 compares the learning experiences between VR and control groups. On average, the students reflected a significantly higher level of satisfaction in using the control format of instructional material compared to using the VR format material.

Table 9: Learning experience: VR vs control groups.

|  | I liked learning using the instructional material format I received | I learned the necessary material using the instructional format I received |
|---|---|---|
| U stat | 24.5 | 58 |
| p-value | .001** | .19 |

** p-value <0.01

Overall, the VR group successfully completed the authentic drilling task at a higher rate than the control group (75% vs 64%). While not statistically significant, the VR group also committed less severe mistakes (1.75 vs 2.14), asked less questions per capita (0.92 vs 1.5), and spent less time (17.67 mins vs 21.57 mins). The intuition and experience towards physical equipment mechanical engineering students have may have also impacted the learning. Future work can study this system's impact on novice learners.

Compared to the control group, the VR group also experienced a larger gain in cognitive assessments. They experienced 14%, 6%, 12%, more gain in the three cognitive constructs, and 8% in total overall scores. Such impact on machining learning outcome could be attributed to the two main features of VR simulation: providing a low-risk learning environment and visual and spatial representation. Common positive feedback from participants include "[VR] allowed me to have a visual representation of the machine and get used[sic] to how to run it" (four participants) and "VR allows for [hands on] practice with the machines in a low risk environment, without fear of accidentally breaking something" (three participants).

## 5 DISCUSSION

In this study, we aimed to explore the extent to which the use of open-ended VR environments can enhance a student's performance and user experience.

*Understanding and self-efficacy*: Despite experiencing greater gains in their cognitive understanding of the material, the VR participants reported less self-efficacy to understand the material. And despite performing better in the psychomotor assessments, the VR group rated their own performance 18% worse than the control group. Conversely, the control group demonstrated less understanding and performed worse, but reported greater self-efficacy. This result is consistent with the well-known cognitive bias studied by Dunning and Kruger: people with lower competence consistently overestimate their ability, and vice versa [47].

*Psychomotor performance and demand:* Despite performing better in the psychomotor assessments, participants who trained in VR felt the physical drilling task was more demanding (20% more mental demand, 8% more physical demand, and 5% more effort compared to control group). The increased demand may partly be due to the increased cognitive load of transferring skills learned in a VR environment to the real-world [48]. To harmonize greater demand with better performance, we hypothesize that because the VR group already had the opportunity to practice in an open-ended environment, they were exposed to more task complexity. One participant shared "it did allow me to try things without fear of consequences." They now had the opportunity to think on how to avoid making the errors they may have encountered in the simulation. One VR participant shared they were "nervous of forgetting small things" during the physical drilling task. The challenge of learning complexity that leads to robust learning is also consistent with our timing measurements. While both groups were given up to 50 minutes for the training portion, most students in the control group took about half that time, whereas many students in the VR group took almost the entire allotted time. Yet, in the actual physical drilling task, the VR group reported less temporal demand and completed in 22% less time.

*Learner satisfaction*: Participants were also more satisfied toward using the control format learning than the VR format. This may partly be explained by issues with the usability of the VR system. Participants pointed out technical bugs in the VR simulation: "the vise handle[sic] and the table y-axis handle[sic] would interfere with each other"; "a black – on the instruction wall when I was supposed to center drill"; "using the chuck key in VR felt a bit glitchy"; and "sometimes it was hard to see the VR text or video". Using VR also required participants to get used to a new tool for learning: "there was a bit of a learning curve for the VR system that was a bit frustrating"; "Learning / getting familiar with it took time". The physical discomfort of using the VR headset was also referred to by the participants: "it was very [tight] and gave me a headache"; "the headset was uncomfortable." These struggles might have put an emotional and psychological toll on participants' learning and practicing experience, and consequently influenced the affective assessments.

*Open-endedness:* Despite our simulation not being fully open-ended, participants still used language that suggests a perception that their learning experience was open-ended. Users pointed out the opportunity to practice: "the VR allows for practice with the machines in a low-risk environment." Other participants reflected on the opportunity to "try things," "get used[sic] to run it," and "practical familiarity." Conversely, no participants described they felt the training was constraining or scripted. One interesting question to explore in the future is what level of open-endedness leads to the perception of open-endedness.

It is interesting to explore the work that would be required to make a fully open-ended simulation. For example, our simulation required users to first install the chuck into the spindle before loosening the chuck jaws, installing an edge finder, and tightening the chuck jaws. In reality, a user can also directly take the chuck from the tool board, install the edge finder in hand, and then install the chuck+edge finder combo into the spindle. The same is true about the deburring task. The current simulation only allows users to debur the top face of the block when it is fixed in the vise. A user in an actual machining setting can directly pick up the block and debur either side. These examples highlight a major distinction. Some open-endedness can be attained by allowing for different pathways through the same tasks performed in the same way. We have developed this form of open-endedness in our paper. However, some open-endedness requires development of *alternative*

*interactions*. The interactions that need to be developed would not create new tasks but *additional* ways of completing the same tasks.

## 6  DESIGN IMPLICATIONS

Designing open-ended virtual reality (VR) training systems presents both limitations and design opportunities that have the potential to shape the future of training of hands-on skills. In this section, we discuss some of the big takeaways of our work and how the design implications from our study can inform the design of open-ended training systems in VR.

*Determining the appropriate level of open-endedness:* While our study results showed the open-ended system for VR training led to a higher completion rate and committed fewer severe mistakes, identifying the right amount of open-endedness and the right moments for open-ended exploration is the key to designing effective open-ended training systems in VR. Striking a balance between providing learners with the freedom to explore and experiment while maintaining a structured learning environment is important. Furthermore, identifying the instances when the system's open-endedness provides opportunities for learning versus hinders the task completion due to too many open-ended pathways is essential. Designers thus need to weigh the cost of adding additional interactions to increase open-endedness, with the benefit of additional learning impact, and strike a balance that allows for open-ended exploration within a defined framework.

*Identifying the inter-dependencies of sub-tasks to design for open-endedness:* Unlike linear path simulations that dictate specific actions, open-ended simulations necessitate identifying multiple rules, constraints, and dependencies within the training system. We observed that designing these open-ended systems can be tedious as they require multiple rounds of testing and input from experts and instructors. This is because the rules and dependencies of the simulation may not be captured in a traditional task analysis, which relies on observing a single instance of task performance. The design process for open-ended simulations should ideally include extensive testing to account for the various potential scenarios and cases that can arise due to the introduced rules and dependencies. By incorporating iterative testing and feedback loops with domain experts, designers can ensure that the open-ended simulation allows for diverse approaches and fosters exploration while still maintaining a structured framework for effective learning experience.

*Using open-endedness to support learners' self-efficacy:* The surprising results of the study revealed that despite performing better in cognitive and psychomotor assessments, participants in the VR training group reported lower self-efficacy and rated their own performance worse than the control group. This unexpected result highlights the need for designers of open-ended VR training systems to consider ways to enhance learners' self-efficacy and confidence. Design opportunities may include incorporating feedback mechanisms or scaffolding techniques to support learners' self-assessment and foster a positive perception of their performance.

*Adapting open-endedness to personalize cognitive learning:* Another important design consideration may be to adapt the open-endedness to the learner's skill levels, prior experience, or familiarity with the tools. Novice learners who are less familiar with the tools and tasks may benefit from a more structured and guided approach initially. By providing a gradual introduction to open-ended options and allowing learners to gain familiarity with the tools and their functionalities, designers can support novices in building a solid foundation of knowledge and skills. As learners progress and become more proficient, the system can gradually offer increased freedom to explore different approaches and solutions. This adaptive approach ensures that learners' open-ended options align with their skill level and promotes a personalized learning experience.

*Scaffolding the learners to support exploration:* Using the features in VR, designers can scaffold the learner's experience to promote explorative learning. Designers can incorporate features that provide feedback and guidance on alternate ways to complete tasks. For example, the system could offer suggestions or demonstrate different approaches to problem-solving, highlighting the benefits and trade-offs of each option. This scaffolding not only enhances learners' understanding of the task but also encourages them to think critically and creatively about the problem at hand. As learners gain a holistic understanding of the task and self-explore the solutions, the system can gradually reduce the nudges to prompt the learners to explore alternative solutions.

## 7  LIMITATIONS AND FUTURE RESEARCH

There were several limitations to our study. First, our study participants consisted of a small sample (26) of engineering students, and do not represent the general population of novice learners. In fact, the intuition and experience towards physical equipment these students came in with may even have enabled them to "connect the dots" and softened the potential differences in learning gains between the different interventions. Future work can study this system's impact on novice learners. Second, our user study compared training in an open-ended simulation with traditional video- and paper-based, direct instruction. Conducting a similar study between open-ended and close-ended VR simulations will allow us to further delineate the impact of open-ended training in simulations. Finally, the instructions within the simulation first guided users along one successful pathway, and then when learners were in the open-ended environment the simulation gave guidance in the form of feedback on errors. We are interested in exploring further scaffolding the experience so users can understand the different equally good alternative choices available to them at any point and their implications.

## 8  CONCLUSION

This paper explores an open-ended approach to the system design for learning psychomotor skills. Our system allowed for multiple pathways to achieve the goal to mimic the open-ended nature of manufacturing tasks. We worked with instructors to define learning goals, perform task-analysis, and identify common mistakes, so users can explore multiple ways for task execution and goal adjustment. To study the impact of our open-ended immersive VR training, we conducted a user study with 26 engineering students. Our results reveal positive trends in favor of our open-ended VR system, with the VR training group successfully completing the machining task of drilling at a higher rate, with fewer mistakes, and in less time compared to the control group. We derived design implications towards the appropriate level of open-endedness, inter-dependencies of sub-tasks, scaffolding, and affective and cognitive gains.

## ACKNOWLEDGMENTS

This work is supported by MIT Abdul Latif Jameel World Education Lab and the d'Arbeloff Fund for Excellence in Education. The authors wish to thank Daniel Gilbert, John Hart, Gregory Osborne, Josh Ramos, and Wade Warman for their help.